\documentclass[conference,a4paper]{IEEEtran}
\IEEEoverridecommandlockouts

\usepackage[utf8]{inputenc}
\usepackage{tikz,pgfplots}
\usetikzlibrary{calc}
\usetikzlibrary{arrows}
\usetikzlibrary{decorations}
\pgfplotsset{compat=newest}
\pgfplotsset{plot coordinates/math parser=false}
\usetikzlibrary{plotmarks}
\usepgfplotslibrary{patchplots}
\usepackage{flushend}
\usepackage{sansmath}
\usepackage{soul}
\usepackage{url}
\newlength\figureheight
\newlength\figurewidth
\usepackage{cite}
\usepackage{amssymb}
\usepackage[cmex10]{amsmath}
\usepackage{algorithm}
\usepackage{algorithmicx,algpseudocode}
\algrenewcommand{\algorithmiccomment}[1]{\hfill \hskip3em$\#$ #1}
\algrenewcommand{\algorithmicreturn}[1]{\State{\bf return} #1}
\usepackage{mdwmath}
\usepackage{mdwtab}
\usepackage[caption=false,font=footnotesize]{subfig}
\usepackage{fixltx2e}
\usepackage{amsthm}
\usepackage{bm}
\usepackage{nicefrac}
\usepackage{mathtools}
\usepackage{placeins}

\newcommand{\ssection}[1]{\subsection[#1]{\centering #1}}

\DeclareMathOperator*{\argmax}{\arg\!\max}

\newtheoremstyle{lem}
  {2ex}
  {2ex}
  {\normalfont\itshape}
  {}
  {\normalfont\bfseries}
  {\newline}
  { }
  {\thmname{#1}\thmnumber{ #2} \thmnote{#3}}

\theoremstyle{lem}

\newcommand{\CC}{\mathbb C}

\renewcommand{\vec}[1]{\bm{#1}}

\renewcommand{\hat}{\widehat}

\newcommand{\evalmap}{\varphi}
\newcommand{\JacobiEval}{\vec J_{\evalmap}}

\begin{document}

\title{Guruswami--Sudan List Decoding for \\Complex Reed--Solomon Codes}

\author{\IEEEauthorblockN{Mostafa H. Mohamed, Sven Puchinger, and Martin Bossert}
\IEEEauthorblockA{Institute of Communications Engineering\\
Ulm University\\
89081 Ulm, Germany\\
\{mostafa.h.mohamed, sven.puchinger, martin.bossert\}@uni-ulm.de}
}

\maketitle

\begin{abstract}
We analyze the Guruswami--Sudan list decoding algorithm for Reed--Solomon codes over the complex field
 for sparse recovery in Compressed Sensing.
We propose methods of stabilizing both the interpolation and the root-finding steps against numerical instabilities, where the latter is the most sensitive.
For this purpose, we modify the Roth--Ruckenstein algorithm and propose
a method to refine its result using Newton's method.
The overall decoding performance is then further improved using
Generalized Minimum Distance decoding based on intrinsic soft information.
This method also allows to obtain a unique solution of the recovery problem.
The approach is numerically evaluated and shown to improve upon recently proposed decoding techniques.
\end{abstract}

\IEEEpeerreviewmaketitle

\section{Introduction} \label{s:Intro}

\noindent
We consider Complex Reed--Solomon (CRS) codes for the application of Compressed Sensing (CS) sparse recovery.
They are first introduced by Wolf \cite{wolf_redundancy_1983} and Marshall \cite{marshall_coding_1984}.
These codes are investigated for a variety of applications in~\cite{henkel1989decodierung,marvasti_efficient_1999,abdelkefi_use_2008,parvaresh_explicit_2008}.

Since they are defined over the complex field, numerical inaccuracies and floating point errors arise when applying any recovery or decoding algorithm.
In \cite{MRZB_2015}, known algebraic decoding algorithms such as the Berlekamp--Massey Algorithm (BMA) \cite[Ch. 7]{berlekamp_1968} and concepts such as Power Decoding (PD)~\cite{schmidt_2010} are adapted to overcome such inaccuracies.

For list decoding algorithms, the Coppersmith--Sudan algorithm \cite{coppersmith_reconstructing_2003} is analyzed in \cite{parvaresh_explicit_2008}.
In the same paper, the Guruswami--Sudan (GS) algorithm \cite{GS} is mentioned, however, not investigated. 
This is due to the presence of the root-finding step which is usually done using the Roth--Ruckenstein (RR) algorithm~\cite{RothRuck2000}, \cite[Page 284]{RothBook}.
Unfortunately, the RR algorithm is unstable when dealing with complex valued numbers.
Within the presented contribution, we use the Newton method \cite[Section 5.6]{Numerik} to achieve a correct decoding result for the GS algorithm.
Generalized Minimum Distance (GMD) decoding \cite{Forney1966} is then used to improve results using intrinsic soft information and to get a unique solution of the recovery problem. 

The paper is organized as follows: In Section \ref{s:Pre}, the definition of CRS codes, notation and basic concepts are established.
All the steps of the GS algorithm, modifications and the use of the Newton method are explained in Section \ref{s:GS}.
In Section \ref{s:GMD_GS}, the proposed GS based GMD decoder is introduced.
Afterwards, its performance using the numerical results are shown in Section~\ref{s:Results}.

\section{Preliminaries and definitions} \label{s:Pre}
\noindent
Let $n$ and $k$ be positive integers fulfilling $k<n$.
The CRS code $\mathcal{CRS}(n,k,d)$ of length $n$ and dimension $k$ is the set
{\small
\begin{equation*}
    \left\{\left. \frac{1}{\sqrt{n}} \left( \vec C\left(\alpha^1\right),\ldots, \vec C\left(\alpha^{n}\right) \right)
        \right| \vec C\left(x\right) \in \CC[x] \land \deg \vec C\left(x\right) < k \right\},
\end{equation*} 
}where $\alpha=e^{-\textrm{j} \frac{2\pi}{n}}$ and $\textrm{j}=\sqrt{-1}$.
The minimum distance of the code is $d=n-k+1$.

The parity-check matrix of a CRS code is denoted as $\vec H$ with dimensions $(n-k)\times n$ and is defined such that
\begin{equation}
\vec{Hc}^T=\vec{0},\qquad \vec c\in \mathcal{CRS}. \label{eq:ParCheckMat}
\end{equation}

Let $\vec H$ be considered as a sensing matrix and $\vec e \in \CC^n$ a sparse vector which can be compressed into $\vec b$ by the
following equation:
\begin{equation}
\vec b^T=\vec{He}^T,\qquad \vec b \in \CC^{n-k},
\end{equation}
where $\vec b$ is the syndrome (in compressed sensing, it is called measurement vector).
The goal is to reconstruct the sparse vector $\vec e$ from the syndrome $\vec b$ using the GS algorithm,
which is interpolation-based and does not use the syndrome directly like in the BMA.
The vector $\vec{r} = \vec b\vec{H}^* + \boldsymbol{\eta}\in  \CC^n$ contains the interpolation points needed by the GS algorithm. 
The term $\vec \eta$ is Gaussian-distributed noise which represents quantization errors, measurement noise and the finite precision of the computation.

Based on the definition of the parity check matrix $\vec H$, the vector $\vec r$ can be considered as some erroneous codeword  with $\vec r = \vec c +\vec e +\vec \eta$. 
Error coordinates in which $\vec e$ is non-zero are denoted by $E =\{i \vert r_i \ne c_i\}$.
The number of errors (or sparsity) is $t=\# E$, where $\#$ gives the cardinality.
Let $\vec \Lambda(x)$ be the error locator polynomial, such that:
\begin{equation}\label{eq:ErrLocLambda}
\vec \Lambda(x) = \prod_{i \in E} (x-\alpha^i).
\end{equation}

To get $\vec \Lambda(x)$, one inputs $\vec r$ to the BMA. If the number of errors is below half the minimum distance $\tau_{BMA} = (d-1)/2$, the algorithm is successful and the non-zero elements in $\vec e$ are calculated using the Gorenstein--Zierler (GZ) algorithm \cite{GZ_1961}.

However, if the number of errors is expected to be beyond $\tau_{BMA}$, list decoders are required since they have a larger decoding radius.
They are called list decoders since the do not usually output one unique solution, but rather a possible list of solutions.

\section{Guruswami--Sudan for CRS codes}\label{s:GS}
\noindent
The GS algorithm is an interpolation-based decoding algorithm that first appeared in \cite{GS}. It can be considered an extension to the Sudan algorithm \cite{Sudan97} where multiplicities were not yet introduced to the interpolation.

The main objective of the algorithm is to find a bivariate polynomial $\vec Q(x,y)$ such that:
\begin{equation}
\vec Q(x,y)=\vec Q_0(x)+\vec Q_1(x)y+\vec Q_2(x)y^2+\dots+\vec Q_\ell(x)y^\ell,
\end{equation}
while satisfying the following conditions:
\begin{enumerate}
\item $\vec Q(\alpha^i,r_i)=0$ with multiplicity $s$, $\forall i=1,\dots,n$
\item $\vec Q(x,y) \neq 0$
\item $\deg \vec Q_i(x) \le s(n-\tau_{GS})-i(k-1)-1$, $\forall i=0,\dots,\ell$
\end{enumerate}

where $\ell$ is the list size, $s$ is the multiplicity and $\tau_{GS}$ is the maximum decoding radius of the GS algorithm and can be calculated using the following relation:
{\small\begin{equation} \label{eqn:GS}
\tau_{GS} < \min \left(\frac{n(2\ell-s+1)}{2(\ell+1)}-\frac{\ell(k-1)}{2s},n-\frac{\ell(k-1)}{s}\right)
\end{equation}
}
By choosing $s$ and $\ell$ large enough, the Johnson radius $\tau_{GS}$ can be achieved, which is given as
\begin{equation}
\tau_{GS} < n- \sqrt{n(n-d)}
\end{equation}

It is shown in \cite[Lemma 4]{GS} that if the three conditions mentioned above are satisfied while having $t\le \tau_{GS}$, then $(y-\vec C(x))$ is a factor of $\vec Q(x,y)$.
The polynomial $\vec C(x)$ is the Inverse Discrete Fourier Transform of the polynomial form of the codeword $\vec c$.
Applying the Discrete Fourier Transform (DFT) on $\vec C(x)$ gives us $\vec c$, thus allowing the recovery of $\vec e$.
Therefore, a root-finding algorithm is used to extract all $y$-roots of $\vec Q(x,y)$.
Note that $\vec Q(x,y)$ can have multiple $y$-roots satisfying the degree constraint of $\vec C (x)$.
As a root-finding algorithm, the RR is considered as it is known to be efficient.

In theory, the decoding process is applicable to CRS codes. 
However, due to the use of complex valued numbers, the stability of the interpolation and root-finding steps comes into question.
Decoding failures are a result of the numerical inaccuracies arising from finite precision of floating point calculations in software or hardware implementations.
In some algorithms, like the RR, comparisons to exact zeros are needed.
Instead, any value smaller than a threshold $\epsilon$ would be considered as a zero. 
In the following subsections, we discuss the stability and accuracy of each step in the GS algorithm.

\ssection{Interpolation}\label{ss:Interpolation}

\noindent
The interpolation step can be considered as the most stable step in the algorithm. 
It consists of solving a linear system of equations for a set of unknowns, the coefficients of the $\ell$ polynomials $\vec Q_i(x)$ for $i=1,\dots,n$ with $s(n-\tau_{GS})-i(k-1)$ coefficients each.

From the aforementioned conditions, condition $1)$ provides the equations needed to get the unknowns.
A linear system of equations can be built and solved using Singular Value Decomposition (SVD).
Since SVD is considered a stable method \cite[Section 4.7]{Numerik}, \cite[Section 2.5]{MatrixComp}, the interpolation step provides a minimal contribution to numerical inaccuracies.

In \cite{Alekhnovich2005Linear,Trifonov2007,BeelenBrandner2010,ZehGentnerAugot-FIAGuruswamiSudan_2011}, the interpolation is done more efficiently using other dedicated methods.
However, the numerical stability of these methods over $\CC$ has not been investigated yet.

\subsubsection*{Erasures}
In literature, there are many different definitions for an erasure.
In this paper, an erasure is defined as a coordinate in the vector $\vec r$ which is (with high probability) erroneous.
In GS interpolation-based decoding, an erased coordinate is simply not used in the interpolation process, such that it has no effect on the output polynomial $\vec Q(x,y)$.
The interpolation procedure is denoted as $GS(\vec r,  \mathcal{I})$, where the set $ \mathcal{I}$ contains the index of the erased coordinates.

\ssection{Root finding}\label{ss:RR}

\noindent
The RR algorithm is shown in Algorithm \ref{algo:RR}.
In polynomial rings over a finite field $\mathbb{F}$, the algorithm is considered to be efficient.
When dealing with polynomials with complex valued coefficients we should make a few changes. 
The modified version of the algorithm, mRR, is shown in Algorithm~\ref{algo:mRR}.

The first modification is using a threshold $\epsilon$ before finding the integer $m$.
The second modification is removing the IF condition in line 12 in Algorithm \ref{algo:RR}. 
Since a deviation from the correct solution is almost inevitable (due to numerical inaccuracies), it does not make sense to keep it. 
As a result, more polynomials are allowed in the set $U$. 
The set will be refined in a later step.

The algorithm tries to find the polynomial $\vec g(x)=g_0 + +\dots+g_{k-1} x^{k-1}$ by finding its coefficients one by one.
Starting from $g_0$, it uses every calculated coefficient to get the next, until all coefficients are calculated.
Assuming there was a small insignificant error in $g_0$, this error propagates with every new coefficient calculated and increases exponentially.
This results in a catastrophic behavior in the high order coefficients.

Another part where things are most likely to go wrong is the first step in the algorithm, where an integer $m$ should be found such that $x^m$ divides $\tilde{\vec Q}(x,y)$.
This translates to checking if all $\ell$ polynomials $\tilde{\vec Q}_i(x)$ for $i=1,\dots,\ell$ are divisible by $x^m$, hence, the first $m$ coefficients are zero.
If in a certain $i-$th step $m$ is wrongly calculated, this will damage not only the $i-$th coefficient but also all the following coefficients after it.

\begin{algorithm}
\caption{Root-finding algorithm (RR) \cite{RothRuck2000,RothBook}}
\label{algo:RR}
{\bf Input:} Bivariate polynomial $\vec Q(x,y)$, dimension $k$, and $\lambda \in \mathbb{N}$ \\
{\bf Global Variables:}  Set $\mathcal{U} \subseteq F_k[x]$\\
\phantom{\bf Global Variables:} Polynomial $\vec g(x) \in  F_k[x]$
\begin{algorithmic}[1]
\If {$\lambda = 0$}
\State $\mathcal{U} = \emptyset$
\EndIf

\State $m \gets$ largest integer such that $x^m$ divides $\vec Q(x,y)$
\State $\vec T(x,y) \gets x^{-m} \vec Q(x,y)$
\State $Z \gets$ set of all distinct $y$-roots of $\vec T(0,y)$ in $F$
\For {each $\gamma \in Z$}
\State $g_\lambda \gets \gamma$
\If {$\lambda < k-1$ }
\State $RR(\vec T(x,xy+\gamma),k,\lambda+1)$
\Else
\If {$\vec Q(x,\vec g(x)) = 0$}
\State $\mathcal{U} \gets \mathcal{U} \cup \{\vec g(x)\}  $
\EndIf
\EndIf
\EndFor 
\end{algorithmic}
\end{algorithm}

To summarize we introduce the following notation: Assume $\hat{\vec C}(x)$ is the inaccurate solution such that $\Delta_i=\hat{C}_i-C_i$.
It can be established that $\Delta_v < \Delta_w$ $\forall v<w$ for $v,w=1,\dots,k$.
As a results of these instabilities, one can not use the mRR algorithm on its own.

\ssection{Newtons method}\label{ss:Newton}

\noindent
Knowing that the root-finding process is not robust enough against small inaccuracies and errors, we use Newton's method in order to find the correct $y$-roots of the interpolation polynomial $\vec Q(x,y)$.
The method is applied to all polynomials found in the set $\mathcal{U}$ outputted from the mRR algorithm.
Those roots $\vec g(x)$ must satisfy
\begin{align*}
\vec Q(\alpha^i, \vec g(\alpha^i)) = 0 \quad \forall i=1,\dots,n.
\end{align*}
Thus, we can find such roots by considering this evaluation map as a function\footnote{Here and in the following, we interpret $\vec g$ both as a vector in $\CC^k$ and a polynomial $\vec g(x)$ of degree $<k$.} in the coefficients $g_0,\dots,g_{k-1}$ of $\vec g(x)$,
\begin{align*}
\evalmap : \CC^k &\to \CC^n, \\
\vec g :=
\begin{bmatrix}
g_0 \\
\vdots \\
g_{k-1}
\end{bmatrix}
&\mapsto
\begin{bmatrix}
\evalmap_1(\vec g) \\
\vdots \\
\evalmap_n(\vec g)
\end{bmatrix}
=
\begin{bmatrix}
\vec Q(\alpha^1,\vec g(\alpha^1)) \\
\vdots \\
\vec Q(\alpha^n,\vec g(\alpha^n))
\end{bmatrix}, 
\end{align*}
and then solving the non-linear system of equations
\begin{align*}
\varphi(\vec g) = \vec 0.
\end{align*}
There are many methods from numerical analysis for solving such non-linear systems approximately. In this paper, we describe how to solve it using Newton's methods for multivariate complex functions $\CC^k \to \CC^n$ \cite{Numerik}.
In Newton's algorithm, we start with a starting point $\vec z_0 \in \CC^k$ and try to get closer to an actual solution $\vec z \in \CC^k$ of $\evalmap(\vec z) = \vec 0$, by an iteration $\vec z_{i-1} \mapsto \vec z_i$.
The iteration is given by solving the linear system of equations
\begin{align*}
(\vec z_i - \vec z_{i-1}) \cdot \JacobiEval(\vec z_{i-1}) = -\evalmap(\vec z_{i-1})
\end{align*}
for the indeterminate $(\vec z_i - \vec z_{i-1})$ and then adding it to $\vec z_{i-1}$, where $\JacobiEval(\vec z_{i-1})$ is the Jacobi matrix of $\evalmap$ at the point $\vec z_{i-1}$,
\begin{align*}
\JacobiEval(\vec z_{i-1}) = 
\begin{bmatrix}
\frac{\partial \evalmap_1}{\partial g_0}(\vec z_{i-1}) & \dots & \frac{\partial \evalmap_1}{\partial g_{k-1}}(\vec z_{i-1}) \\
\vdots & \ddots & \vdots \\
\frac{\partial \evalmap_n}{\partial g_0}(\vec z_{i-1}) & \dots & \frac{\partial \evalmap_n}{\partial g_{k-1}}(\vec z_{i-1}) \\
\end{bmatrix}
\in \CC^{n \times k}.
\end{align*}

\begin{algorithm}
\caption{Modified root-finding algorithm (mRR)}
\label{algo:mRR}
{\bf Input:} Bivariate polynomial $\vec Q(x,y)$, dimension $k$, and $\lambda \in \mathbb{N}$ \\
{\bf Global Variables:}  Set $\mathcal{U} \subseteq \CC_k[x]$\\
\phantom{\bf Global Variables:} Polynomial $\vec g(x) \in  \CC_k[x]$
\begin{algorithmic}[1]
\If {$\lambda = 0$}
\State $\mathcal{U} = \emptyset$
\EndIf
\textcolor{red}{
\State $\tilde{\vec Q}(x,y) \gets \vec Q(x,y)$
\If {$\tilde{\vec Q}_{i,j} < \epsilon$}
\State $\tilde{\vec Q}_{i,j} \gets 0$
\EndIf
}

\State $m \gets$ largest integer such that $x^m$ divides \textcolor{red}{$\tilde{\vec Q}(x,y)$}
\State $\vec T(x,y) \gets x^{-m} \textcolor{red}{\tilde{\vec Q}(x,y)}$
\State $Z \gets$ set of all distinct $y$-roots of $\vec T(0,y)$ in $\CC$
\For {each $\gamma \in Z$}
\State $g_\lambda \gets \gamma$
\If {$\lambda < k-1$}
\State $RR(\vec T(x,xy+\gamma),k,\lambda+1)$
\Else
\State \textcolor{red}{$\mathcal{U} \gets \mathcal{U} \cup \{\vec g(x)\}$}
\EndIf
\EndFor 
\end{algorithmic}
\end{algorithm}

The exact expression of $\evalmap_i(\vec g)$ for $i=1,\dots,n$ is
\begin{align*}
\evalmap_i(\vec g) &= \sum_{\mu} \sum_{\nu} Q_{\mu,\nu} (\alpha^i)^\mu (\vec g(\alpha^i))^\nu \\
&= \sum_{\mu} \sum_{\nu} Q_{\mu,\nu} \alpha^{i\mu} \left(\sum_{\xi=0}^{k-1} g_\xi \alpha^{i\xi}\right)^\nu,
\end{align*}
where the indices $\mu,\nu$ run over the degree restrictions given by the GS interpolations problem. Thus,
\begin{align*}
\frac{\partial \evalmap_i}{\partial g_j}(\vec g) = \sum_{\mu} \sum_{\nu} Q_{\mu,\nu} \nu \alpha^{i(\mu+j)} \left(\sum_{\xi=0}^{k-1} g_\xi \alpha^{i\xi}\right)^{\nu-1},
\end{align*}
which gives us an explicit description of the Jacobi matrix $\JacobiEval(\vec g)$ for any $\vec g \in \CC^k$.

\subsubsection*{Faster Convergence}
It is important to note that Newton's method does not always converge. 
However, if it does, it often locally converges quadratically in the number of iterations. 
In order to achieve this, a ``good'' starting point $\vec z_0$ must be chosen.
As discussed at the end of Section \ref{ss:RR}, the roots provided by the mRR algorithm contains inaccuracies in its coefficients that increases with the degree of its monomials at an exponential rate.
Assuming mRR provided us with the polynomial $\vec p=p_0+\dots+p_{k-1} x^{k-1}$, it turns out that choosing the $\vec z_0= p_0+\dots+p_m x^{m}$, where $m<k-1$ provides a faster rate of convergence.
In this paper, we chose $m=\lfloor k/2 \rfloor$ and in Section \ref{s:Results} it is shown that it serves as a ``good'' starting point since it is very close to an actual $y$-root of $\vec Q(x,y)$.

At this point, a list of roots for $\vec Q(x,y)$ has been obtained and in terms of list decoding the job is already done.
Although results obtained show an output list of size one with high probability, this is not always guaranteed.
A procedure is needed to provide a single solution to reconstruct the sparse vector $\vec e$.
This is done with the aid of the GMD concept, where multiple error/erasure decoding trials using GS take place with the help of soft information.

\section{Guruswami--Sudan based Generalized Minimum Distance decoding}\label{s:GMD_GS}
\noindent
GMD was introduced by Forney in \cite{Forney1966}.
The basic idea behind GMD is using an error/erasure decoder for a number of decoding trials by exploiting soft information.
In each trial, an increasing number of least reliable positions is erased.
The GS algorithm can be seen as an error/erasure decoder.
Erased coordinates are simply points that are not considered in the interpolation step.

In the case of CRS codes, the soft information can be found by simply decoding the received word using any algebraic decoding algorithm \cite[Section 7.3]{ZoerleinDiss}.
In this paper, the BMA is used as a first decoding step.
If it provides a proper error locator polynomial $\vec \Lambda(x)$, the sparse error vector is calculated using the Gorenstein--Zierler algorithm \cite{GZ_1961}.
In case it failed, its output is used as soft information for the GS algorithm.
The proposed algorithm is shown in Algorithm \ref{algo:GS_GMD}.

A single trial of decoding using the GS algorithm provides a list of possible solutions. 
Since sparse recovery requires a single output, we do more decoding trials.
For each trial, the number of erased positions $\rho$ (points not considered in interpolation) is increased.
Since the number of interpolation points change, the parameters $\ell$ and $s$ need to be recalculated.
They are chosen such that they are as minimum as possible and satisfy Equation \eqref{eqn:GS}.
The output of the GS is the list $\mathcal{U}$.
Each element in this list is input to the Newton method and checked to see if it provides a sparse error vector which falls in our designed decoding radius $\tau$.
Vectors that pass this check are then saved in the list $\mathcal{L}$.
Each entry in $\mathcal{L}$ gets a score depending on how many times it appeared.
In the end of the algorithm, the vector that appeared the most is considered to be the solution.

\begin{algorithm}
\caption{GS based GMD decoding}
\label{algo:GS_GMD}
{\bf Input:} Vector $\vec r$, length $n$, dimension $k$ and radius $\tau$\\
{\bf Initialization:} $\mathcal{L} \gets \{\}$, $\mathcal{U} \gets \{\}$ and $\rho \gets 0$ 
\begin{algorithmic}[1]
\State $\vec \Lambda (x) \gets$ BMA$(\vec r)$
\If {$\vec \Lambda (x)$ is a proper error locator} 
\State $\hat{\vec e} \gets $ GZ$(\vec r,\vec \Lambda(x))$
\Return{$\hat{\vec e}$}
\Else
\State $\vec\lambda \gets \vec \Lambda(\alpha^i)$ $\forall i=1,\dots,n$
\Comment{Soft information}
\While{$\rho<\tau$}
\State $\mathcal{I} \gets$ points not to be included in GS (based on $\vec \lambda$)
\State $\tau_{GS} \gets \tau - \rho$
\State $n \gets n-\rho$
\State Choose $\ell$, $s$ such that Equation \eqref{eqn:GS} is satisfied.  
\State $\vec Q(x,y) \gets$ GS$(\vec r,\mathcal{I})$ 
\Comment Section \ref{ss:Interpolation}
\State $\mathcal{U} \gets$ mRR$(\vec Q(x,y),k,0)$ 
\Comment Section \ref{ss:RR}
\For {each $\vec U \in \mathcal{U}$}
\State $\vec W \gets$ Newton$(\vec U)$ 
\Comment Section \ref{ss:Newton}
\State $\vec w \gets DFT(\vec W)$
\State $\tilde{\vec e} \gets \vec r-\vec w$
\If{$supp(\tilde{\vec e}< \epsilon) \le \tau $}
\If{$\tilde{\vec e} \in \mathcal{L}$}
\State $S(\tilde{\vec e}) \gets S(\tilde{\vec e})+1$
\Else 
\Comment New entry in $\mathcal{L}$
\State $\mathcal{L} \gets \mathcal{L} \cup \tilde{\vec e}$
\State $S(\tilde{\vec e}) \gets 1$
\EndIf
\EndIf
\EndFor
\State $\rho \gets \rho+1$ 
\EndWhile
\State $\hat{\vec e} \gets \displaystyle\argmax_{\vec l \in \mathcal{L}}\>S(\vec l)$ 
\Return{$\hat{\vec e}$} 
\EndIf
\end{algorithmic}
\end{algorithm}

\section{Numerical Results}\label{s:Results}
\noindent
The performance of Algorithm \ref{algo:GS_GMD} is evaluated numerically for the cases of a $\mathcal{CRS}(32,8)$ and $\mathcal{CRS}(16,4)$ codes.
It is compared to BMA and PD with Continuity Assisted Decoding (CAD) \cite[Algorithm 7.1]{ZoerleinDiss}. 
The simulations are made for $10000$ samples in a noiseless as well as a noisy environment with noise vector $\vec \eta$. 
The noisy environment has been considered where the real and imaginary parts of the complex-valued noise components are drawn from a normal distributed random source with zero mean and standard deviation $\sigma=\nicefrac{\sigma_{\vec{\eta}}}{\sqrt{2}}$ with $\sigma_{\vec{\eta}}=10^{-7}$.

Subsequently, boxplots (as modeled in~\cite{tukey77}) are used to visualize the distribution of given datasets.
The main part of the boxplot is built by a rectangle, which resembles the values between first and third quartile.
The median corresponds to a black horizontal bar within this box and the mean is given as a circle.

\begin{figure} 
\centering
\setlength\figureheight{0.65\columnwidth} 
\setlength\figurewidth{0.8\columnwidth}
\input{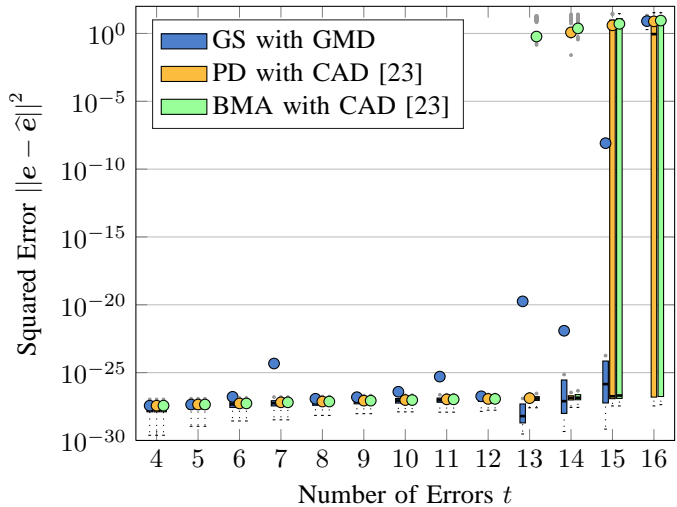} 
\caption{Boxplots illustrating the squared error $\left\vert\left\vert\vec{e}-\hat{\vec{e}}\right\vert\right\vert^2$ for $10000$ samples in noiseless scenario for $\mathcal{CRS}(32,8)$ code for different decoding schemes and number of errors $t$.} \label{fig:Results1}
\end{figure}

The parameters for the $\mathcal{CRS}(32,8)$ code are as follows: GS decoding radius is $\tau_{GS}=15$, half minimum distance $\tau_{BMA}=12$ and power decoding radius $\tau_{PD}=13$.
The results for the simulation for the noiseless and noisy cases are shown in Figures \ref{fig:Results1} and \ref{fig:Results1_noisy} respectively.

\begin{figure}
\centering
\setlength\figureheight{0.65\columnwidth} 
\setlength\figurewidth{0.8\columnwidth}
\input{Results1_noisy.tikz} 
\caption{Boxplots illustrating the squared error $\left\vert\left\vert\vec{e}-\hat{\vec{e}}\right\vert\right\vert^2$ for $10000$ samples in noisy scenario ($\sigma_{\vec{\eta}} = 10^{-7}$) for $\mathcal{CRS}(32,8)$ code for different decoding schemes and number of errors $t$.} \label{fig:Results1_noisy}
\end{figure}

\begin{figure} 
\centering
\setlength\figureheight{0.65\columnwidth} 
\setlength\figurewidth{0.8\columnwidth}
\input{Results2.tikz} 
\caption{Boxplots illustrating the squared error $\left\vert\left\vert\vec{e}-\hat{\vec{e}}\right\vert\right\vert^2$ for $10000$ samples in noiseless scenario for $\mathcal{CRS}(16,4)$ code for different decoding schemes and number of errors $t$.} \label{fig:Results2}
\end{figure}

\begin{figure} 
\centering
\setlength\figureheight{0.65\columnwidth} 
\setlength\figurewidth{0.8\columnwidth}
\input{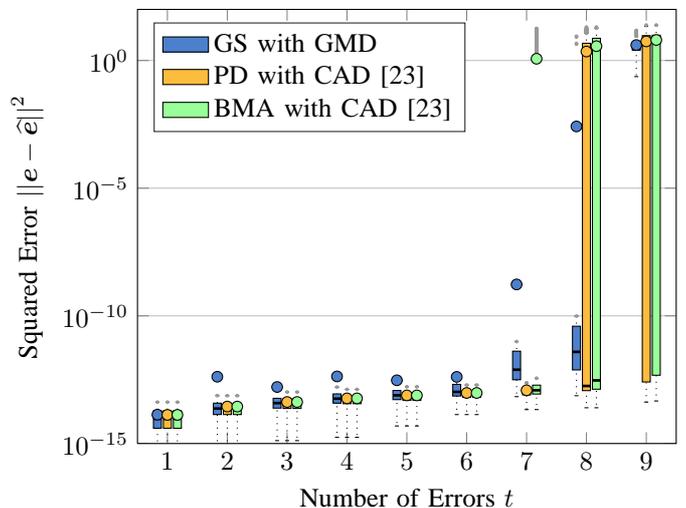}
\caption{Boxplots illustrating the squared error $\left\vert\left\vert\vec{e}-\hat{\vec{e}}\right\vert\right\vert^2$ for $10000$ samples in noisy scenario ($\sigma_{\vec{\eta}} = 10^{-7}$) for $\mathcal{CRS}(16,4)$ code for different decoding schemes and number of errors $t$.} \label{fig:Results2_noisy}
\end{figure}

In the noiseless scenario (Figure \ref{fig:Results1}), the functionality of the proposed algorithm is shown.
With it being able to find the correct sparse error vector with high accuracy almost as good as its competitors, with only of a few insignificant outliers. 
However, since it is based on interpolation, the radius surpasses those of the others.
When noise is added (Figure \ref{fig:Results1_noisy}), the accuracy is lost although the noise level is really small $\sigma_{\vec{\eta}}=10^{-7}$.
This figure shows how highly sensitive is interpolation based decoding to noise.

The same thing can be concluded when using the $\mathcal{CRS}(16,4)$ code. 
The new parameters are as follows: GS decoding radius is $\tau_{GS}=8$, half minimum distance $\tau_{BMA}=6$ and power decoding radius $\tau_{PD}=7$.
The noiseless and noisy cases are shown in Figures \ref{fig:Results2} and \ref{fig:Results2_noisy} respectively.

In Figure \ref{fig:Results2_noisy}, the effect of noise can still be seen, although its impact is not as large as for the $\mathcal{CRS}(32,8)$ code.
Of course the length of the code plays a role the impact of noise. However, the dominant part comes from inaccuracies arising from the root finding step, which increases exponentially with the dimension $k$.

\section{Conclusion}\label{s:Conclusion}

\noindent
The possibility to use the GS algorithm for sparse error recovery in CRS codes has been established.
Aided by the Newton method, inaccurate results produced by the RR root-finding algorithm can be refined and often resulting in an output list size equal to one with high probability.
To get a single solution, GS based GMD decoding is used.
The overall algorithm is able to function properly away from numerical instabilities.
Numerical simulations shows good performance and an increase in the decoding radius when compared to previous results under low noise conditions.
However in a noisy scenario, previous results show more robustness to numerical instability.
The proposed algorithm is sensitive to noise, which is still an issue to be tackled.
Also the effect of cost efficient algorithms and the change of parameters on the performance and stability is still an open question.

\section*{Acknowledgments}

\noindent
The authors would like to thank the anonymous reviewers for their valuable comments and suggestions to improve the quality of the paper.
This work was supported by the German research council Deutsche Forschungsgemeinschaft (DFG) under Grant Bo~867/35-1.

\bibliographystyle{IEEEtran}
\bibliography{SCC2017}

\end{document}